# Open Science and Authorship of Supplementary Material. Evidence from a Research Community.[1]

Andrea Mannocci[*], Ornella Irrera[*,**] and Paolo Manghi[*,†]

[*] *andrea.mannocci@isti.cnr.it; paolo.manghi@isti.cnr.it*
ISTI-CNR, Pisa, Italy

[**] *ornella.irrera@studenti.unipd.it*
University of Padua, Italy

[†] *OpenAIRE AMKE, Athens, Greece*

**Introduction**

While, in early science, most of the papers were authored by a handful of scientists, modern science is characterised by more extensive collaborations, and the average number of authors per article has increased across many disciplines (Baethge, 2008; Cronin, 2001; Fernandes & Monteiro, 2017; Frandsen & Nicolaisen, 2010; Wren et al., 2007). Indeed, in some fields of science (e.g., High Energy Physics), it is not infrequent to encounter hundreds or thousands of authors co-participating in the same piece of research[2]. Such intricate collaboration patterns make it difficult to establish a correct relationship between contributor and scientific contribution and hence get an accurate and fair reward during research evaluation (Brand, Allen, Altman, Hlava, & Scott, 2015; Vasilevsky et al., 2021; Vergoulis et al., 2022). Thus, as widely known, scientific authorship tends to be a rather hot-button topic in academia, as roughly one-fifth of academic disputes among authors stem from this (Dance, 2012).

Open Science, however, has the potential to disrupt such traditional mechanisms by injecting into the "academic market" new kinds of "currency" for credit attribution, merit and impact assessment (Mooney & Newton, 2012; Silvello, 2018). To this end, the new practices of supplementary research data (and software) deposition and citation could be perceived as an opportunity to diversify the attribution portfolio and eventually give credit to the different contributors involved in the diverse phases of the lifecycle within the same research endeavour (Bierer, Crosas, & Pierce, 2017; Brand et al., 2015).

While, on the one hand, it is known that authors' ordering tells little or nothing about authors' roles and contributions (Kosmulski, 2012), on the other hand, we argue that variations of any kind in author sets of paired publications and supplementary material can be indicative. Despite being unclear the actual reason behind such a variation, the presence of a fracture between the publication and research data realms might suggest once more that current practices for research assessment and reward should be revised and updated to capture such peculiarities as well.

---

[1] Work co-funded by the EC H2020 project OpenAIRE-Nexus (Grant agreement: 101017452).
[2] https://dx.doi.org/10.1016/S0375-9474(14)00601-0

In (Mannocci, Irrera, & Manghi, 2022), we argue that modern Open Science Graphs (OSGs) can be used to analyse whether this is the case or not and understand if the opportunity has been seized already. By offering extensive metadata descriptions of both literature, research data, software, and their semantic relations, OSGs constitute a fertile ground to analyse this phenomenon computationally and thus analyse the emergence of significant patterns.

As a preliminary study, in this paper, we conduct a focused analysis on a subset of publications with supplementary material drawn from the European Marine Science[3] (MES) research community. The results are promising and suggest our hypothesis is worth exploring further. Indeed, in 702 cases out of 3,075 (22.83%), there are substantial variations between the authors participating in the publication and the authors participating in the supplementary dataset (or software), thus posing the premises for a longitudinal, large-scale analysis of the phenomenon.

**Data and methods**

*Data*
The present study is possible thanks to the ever-increasing availability of metadata about research products in the last decade and, in particular, to the emergence of Open Science Graphs (OSGs), enabling unrestricted access to graph representations of metadata about people, artefacts, and institutions involved in the research lifecycle. The represented information may span entities such as research products (e.g., publications, research data, software) as well as their content, organisations, researchers, services, projects, funders, and their semantic relations to support stakeholder needs, such as discovery, reuse, reproducibility, statistics, trends, validation, and impact assessment (Aryani, Fenner, Manghi, Mannocci, & Stocker, 2020).

In this analysis, we focus on the European Marine Science (MES) research community extracted from the OpenAIRE Research Graph[4] (Paolo Manghi, Atzori, Bardi, et al., 2020), one of the core products delivered by OpenAIRE AMKE[5], a not-for-profit legal entity operating an infrastructure that offers global services in support of Open Science scholarly workflows. As of December 2021, the whole dataset of research products participating in the MES research community comprehends 104,191 publications, 118,110 datasets[6], and 1,105 software.

*Methods*
First and foremost, we filtered the publications interested in at least one relation with datasets or software regardless of the semantics, leaving us with 4,928 publications, 9,745 datasets, and 28 software. The semantic relations between publications and datasets are 16,123, while the ones between publications and software are 58.

To select a publication $p$ and a supplementary dataset (or a software) $d$ produced within the same research effort (e.g., $p$ describes a research effort to conduct a measuring campaign, contextually producing the dataset $d$), we relied on semantics and selected just the $p \rightarrow d$ *pairs* related by the relation *IsSupplementedBy*, which is the DataCite relation type indicating that $d$ is supplementary material for $p$ (DataCite Metadata Working Group, 2021). This filtering returned 3,211 pairs: 3,188 involve publications and datasets, while 23 involve publications and software.

---

[3] https://mes.openaire.eu
[4] https://graph.openaire.eu
[5] https://www.openaire.eu
[6] In the following, we use the term "dataset" in its most general acceptation as an umbrella concept comprehending proper datasets, such as CSVs files or compressed archives of a file collection, but also figures and tables as they are indistinguishable in our data.

Then, we filtered out unwanted noise accidentally created by the deduplication algorithm operating during the construction of the OpenAIRE Research Graph to reconcile the manifold instances of the same research product (Paolo Manghi, Atzori, De Bonis, & Bardi, 2020). The full-fledged details of the business logic behind the deduplication algorithm go beyond this paper's scope; however, several distinct research datasets were merged in a few cases as they misleadingly shared the same, rather generic title (e.g., "Index data"). In such cases, the information of the distinct instances gets merged into a representative record, and semantic relations are propagated across all the research products participating in the clique. Since this would alter the results of the analysis, we filtered the 136 pairs interested, thus obtaining 3,075 pairs for the analysis, of which 3,052 are involved in the analysis of datasets while 23 in the analysis of software.

Finally, we also processed the author lists of the filtered pairs by selecting (#author_position, full_name) of each author and accumulating them into hash sets, thus preventing unnecessary repetitions.

Once the relevant pairs were selected, we proceeded with analysing the author sets and their interpretation. Let $A_p$ be the set of authors of a publication $p$ and $A_d$ the set of authors of a supplementary dataset (or software) $d$; we manually annotated the pairs with the following *authorship variation events*:
- *Additions*: $A_d$ contains one or more authors that are not present in $A_p$;
- *Removals*: $A_d$ lacks one or more authors that are present in $A_p$;
- *Shuffles*: provided that $|A_p \cap A_d| > 1$, the relative ordering of the authors in $A_p \cap A_d$ is altered. In this way, shuffles do not account for author position changes due to the addition or removal of other authors. Similarly, additions and removals alone do not automatically imply a shuffle.

The observed variations are intended as binary, as the number of events occurring for the single pair is not considered. As an example, $A_p = \{(1, a), (2, b), (3, c)\}$; $A_d = \{(1, c)\}$ counts as one removal, but not as a shuffle, while $A_p = \{(1, a), (2, b), (3, c)\}$; $A_d = \{(1, c), (2, x), (3, a)\}$ would count as one addition, one removal, and a shuffle.

We also flagged *Exceptions* whenever a pair showed anomalies, preventing our annotation task.

**Results**

Focusing on datasets only, 683 pairs out of 3,052 (22.37%) exhibit substantial variations in authorship between the publication and the dataset. In particular, a total of 910 variation events have been marked down: 134 additions (14.73%), 608 removals (66.81%), and 168 shuffles (18.46%). The variation events are possibly co-occurring, as depicted in Figure 1; 213 pairs (31.18%) exhibit at least two authorship variation events at the same time. No authorship variation event was registered for the remaining 2,369 pairs (77.63%). These include 44 problematic pairs that unfortunately prevented us from proceeding with our annotation task. In 22 of these exceptions, for example, $A_d$ attributes the work to a group (e.g., "data curation team") rather than an individual. For 10 others instead, the two author sets have a null intersection and thus are incomparable for reasons often beyond our knowledge. The observed issues, the possible causes, as well as the rest of the data and results, can be openly accessed here[7].

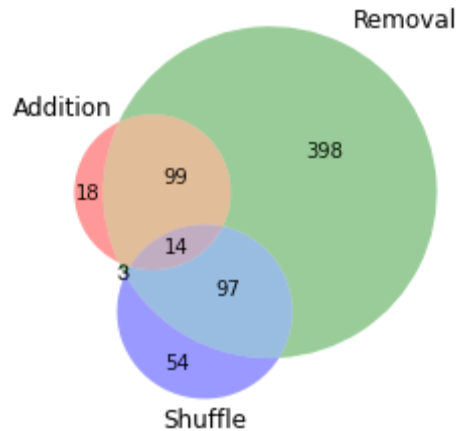

Figure 1 - Observed authorship variation events involving supplementary data

Regarding software, instead, 19 pairs out of 23 (82.6%) exhibit substantial variations in authorship between the publication and the software, and we marked down 24 variation events. A total of 3 pairs (12.5%) were involved with additions, 18 with removals (75%), and 3 (12.5%) with shuffles, as shown in Figure 2. Only 4 pairs (17.4%) exhibit no sort of authorship variation.

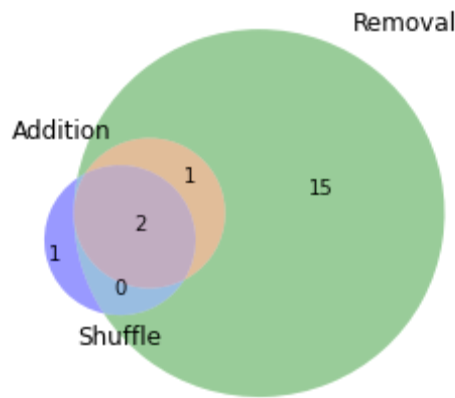

Figure 2 - Observed authorship variation events involving supplementary software

**Discussion**
Overall, the most frequent variation events are removals, while shuffles immediately follow. This could suggest that dataset and software may require skills that not every author in $A_p$ has or that, by any means, some of them have a scarce interest in appearing in the supplementary material. The nature of shuffles is even harder to grasp. While this could be accidental in some cases, we believe that in many occurrences, this is intentional; in fact, 93 shuffles out of 168 (55.35%) involving datasets involve non-adjacent authors, making it harder to happen by mere chance.

---
[7] https://zenodo.org/record/6787089

Additions are the least frequent variation events; nonetheless, this suggests the existence of a considerable "submerged workforce" taking part in research support activities while being underrepresented in literature and hence in traditional rewards and evaluation frameworks.

On a side note, the software has a higher percentage of variation events than datasets, potentially taking to the extreme the aforementioned considerations. Unfortunately, the few data points do not permit us to move further hypothesis.

Generally speaking, a considerable amount of pairs presents substantial variations between the authors participating in the publication and the authors participating in the dataset (or software) deposition, thus indicating the presence within the examined research community of an alternative accreditation scheme in parallel to the traditional one (i.e., same authors in literature and supplementary material). This confirms our initial hypothesis and poses the premises for a longitudinal, large-scale analysis of the phenomenon.

*Limitations*
The present study has a few limitations, which, however, do not affect the validity of the results here described.

First and foremost, there is inherent uncertainty regarding semantic relations specified among literature and research data records. The semantics expressed by the relation is defined by the user (e.g., researcher, librarian, curator) taking care of the deposition process (e.g., on Zenodo). In some cases, the semantics could even be omitted (i.e., both a publication and a dataset are deposited, but the relation is missing). Hence, relations are prone to human errors, as it might not be very straightforward which semantics is the most appropriate. On Zenodo, for example, the choice is drawn from a dropdown menu with scarce or limited guidance on the rationale behind each option.

To mitigate this aspect, we plan to run a heuristic over pairs tied by "vanilla" relations (e.g., *Cites*, *References*) and infer the unintentionally lost relations indicating supplementary material. A viable strategy consists of retrofitting as *IsSupplementedBy* relations all the *Cites* and *References* relations when the author sets share at least an author and the publication date indicates that the two records could be contextual (e.g., within six months apart).

A possible generalisation of the above heuristic would rely on multiple metadata fields such as date of publication, title, subjects, and author set to create a feature vector describing research outputs. Then, to infer additional supplementary material, the representations of known $p{\rightarrow}d$ pairs could be used to train a binary classifier or to compute a confidence interval. If the similarity between two given feature vectors tied by "vanilla" relations lies within a computed confidence interval, then the semantics has been probably misassigned, and thus it can be retrofitted as *IsSupplementedBy*.

A second limitation lies in the comparison of the two author sets, which brings the long-standing challenge of author disambiguation into the picture, as the same author could be spelt out in two different ways in the two related sets. This is crucial, especially from the perspective of a large-scale, longitudinal analysis. Indeed, the presence of ORCID iDs could ease the task. Still, as their coverage is far from ideal, to address the problem, we could rely on the deduplication framework of OpenAIRE (P. Manghi, Mikulicic, & Atzori, 2012; Paolo Manghi, Atzori, De Bonis, et al., 2020) in order to compute the distance between single authors and sets. It is worth noting that, in contrast to the standard deduplication task (i.e., establish the equivalence of similar research products), we are comparing author sets belonging to research outputs different in kind (i.e., literature with non-literature). These sets may not necessarily contain the same authors; hence, the methods provided for the deduplication need to be customised according to the needs of this specific task.

**Future work**
In the future, we intend to expand this study in several directions. First of all, we will repeat the analysis both on a larger scale and in other research communities to identify global trends as well as practices locally specific or shared across communities. In particular, further investigation is required for software, as in MES, we had just a few data points at our disposal.

Then, it would be interesting to unfold the analysis throughout time and understand whether trends associated with authorship variation events emerge. If the number of variations increases over the years, this could indicate that researchers and collaborators are taking more and more advantage of Open Science practices to diversify their portfolio of contributions. Similarly, correlating such trends with the introduction of novel Open Science policies at a national and supranational level would help to understand better if and how the scientific community receives them.

Finally, we plan to detect patterns revealing a possible agency behind authorship variation events. One possible way is to take advantage of novel scholarly services to precisely track author roles and contributions (Vergoulis et al., 2022) to detect if a given researcher systematically takes part in or avoids specific phases of the research lifecycle. Another viable strategy could correlate authorship variation events with authors' seniority. For example, data could suggest that senior staff members are less involved or, by any means, less interested in participating, or getting rewarded, for the production of datasets or software development, thus confirming a bias toward the *status quo* of research assessment.

**Acknowledgements**
This work was partially funded by the EC H2020 project OpenAIRE-Nexus (Grant agreement 101017452).